\documentclass[pdflatex,sn-basic]{sn-jnl}

\usepackage{moreverb}
\usepackage{relsize}
\usepackage{nicefrac}
\usepackage{bm}
\usepackage{amssymb}
\usepackage{amsmath}
\usepackage{setspace}
\usepackage{booktabs}
\usepackage{multirow}
\usepackage{makecell}
\usepackage{dsfont}
\usepackage{bbold}
\usepackage{soul}
\usepackage{color}
\usepackage{caption} 
\usepackage{mathrsfs}
\usepackage{optidef}
\usepackage{float}
\usepackage{comment}
\usepackage{hyperref}
\usepackage{multicol}
\usepackage{algorithm} 
\usepackage{algpseudocode}
\usepackage{actuarialsymbol}
\usepackage{cleveref}
\usepackage{leftidx}
\hypersetup{
    colorlinks,
    linkcolor={red!50!black},
    citecolor={blue!50!black},
    urlcolor={blue!80!black}
}

\newcounter{algsubstate}


\usepackage{comment}
\usepackage{url} 
\usepackage{subcaption}
\usepackage{MnSymbol}%
\usepackage{wasysym}%

\jyear{2022}%

\theoremstyle{thmstyleone}%
\theoremstyle{thmstyletwo}%

\theoremstyle{thmstylethree}%

\begin{document}

\title[A Generalized Framework for Microstructural Optimization]{A Generalized Framework for Microstructural Optimization using Neural Networks}


\author[1]{\fnm{Saketh} \sur{Sridhara}}\email{ssridhara@wisc.edu}
\equalcont{These authors contributed equally to this work.}

\author[1]{\fnm{Aaditya} \sur{Chandrasekhar}}\email{achandrasek3@wisc.edu}
\equalcont{These authors contributed equally to this work.}

\author*[1,2]{\fnm{Krishnan} \sur{Suresh}}\email{ksuresh@wisc.edu}

\affil[1]{\orgdiv{Department of Mechanical Engineering}, \orgname{University of Wisconsin-Madison}, \city{Madison}, \state{WI}, \country{USA}}
\affil[2]{\orgdiv{Department of Engineering Physics}, \orgname{University of Wisconsin-Madison}, \city{Madison}, \state{WI}, \country{USA}}


\abstract{
Microstructures, i.e., architected materials, are designed today, typically, by maximizing an objective, such as bulk modulus, subject to a volume constraint. However, in many applications, it is often more appropriate to impose constraints on other physical quantities of interest.\\ 

In this paper, we consider such generalized microstructural optimization problems where any of the microstructural quantities, namely, bulk, shear, Poisson ratio, or volume, can serve as the objective, while the remaining can serve as constraints. In particular, we propose here a neural-network (NN)  framework to solve such problems. The framework relies on the classic density formulation of microstructural optimization, but the density field is represented through the NN's weights and biases.  \\

The main characteristics of the proposed NN framework are: (1) it supports automatic differentiation, eliminating the need for manual sensitivity derivations, (2) smoothing filters  are not required due to implicit filtering, (3) the framework can be easily extended to multiple-materials, and (4) a high-resolution microstructural topology can be recovered through a simple post-processing step. The framework is illustrated through a variety of microstructural optimization problems.\\
}




\maketitle

\section{Introduction}
\label{sec:intro}
In microstructural optimization, one aims to find the optimal topology, within a representative unit cell, that maximizes the desired property.  This has many applications in engineering; for example,  energy dissipation \cite{asadpoure2017topology}, fluid applications \cite{guest2006optimizing}, thermal applications \cite{zhou2008computational}, phononic applications \cite{sigmund2003systematic}, medical implants \cite{hsieh2021architected}, and so on. Further, with the advent of additive manufacturing, the fabrication of such microstructures is possible today.

In a typical microstructural optimization problem, one attempts to maximize a quantity of interest (such as bulk modulus), subject to a mass constraint (or equivalently, volume-fraction constraint). However, in many applications, the desired mass is not known a priori.  Therefore, instead of imposing an arbitrary mass constraint, we consider imposing constraints on other physical quantities. The main objective of this paper is to develop a framework for solving such generalized microstructural optimization problems.

Further, it has been observed \cite {sigmund1994materials} that the design of negative Poisson ratio (NPR) materials, using standard optimization techniques, can be particularly challenging. Either specialized methods need to be developed \cite{yin2001optimality} or heuristic parameters must be used \cite{xia2015design}. Here, we show that standard L-BFGS optimization can be used for the robust design of NPR materials.

The remainder of this paper is organized as follows. First, the critical concept of homogenization is briefly reviewed in Section \ref{subsec:homogenization}. Then, the current methods of  microstructural optimization are reviewed in Section \ref{subsec:litreview}, with an emphasis on the classic density-based formulation.  Then, in Section \ref{subsec:generalizedProblems}, a generalized microstructural optimization problem. To solve such a class of problems, we introduce a neural-network (NN) framework in Section \ref{sec:method_DesignRepresentationNN} where the density field is represented through the weights associated with the NN. This allows for automatic sensitivity computation which is essential for solving generalized problems.  In Section \ref{subsec:augLag} the methodology to solve such generalized problems is discussed followed by the proposed algorithm  in Section \ref{subsec:algorithm}. Numerical examples are presented in Section \ref{sec:experiments}. Finally, conclusions and open issues are discussed in Section \ref{sec:conclusion}.

\section{Background}
\label{sec:background}

\subsection{Homogenization}
\label{subsec:homogenization}
\begin{figure}[H]
	\begin{center}
    \includegraphics[scale=0.5]{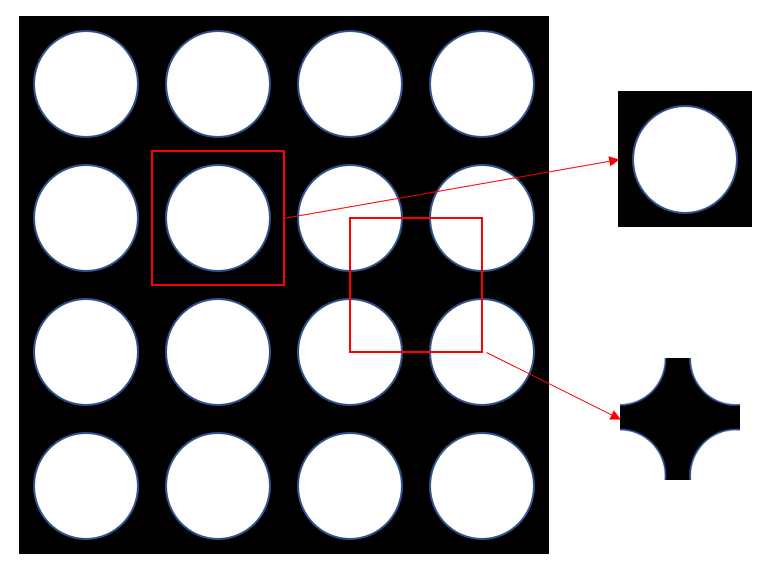}%
		\caption{Locally periodic microstructures and unit cells.}
		\label{fig:UnitCell}
	\end{center}
\end{figure}

In microstructural design, a common hypothesis is that the microstructure is locally periodic, and there is scale separation. For example, Figure \ref{fig:UnitCell} illustrates locally periodic microstructures and two representative unit cells.

\begin{figure}[H]
	\begin{center}
    \includegraphics[scale=0.75]{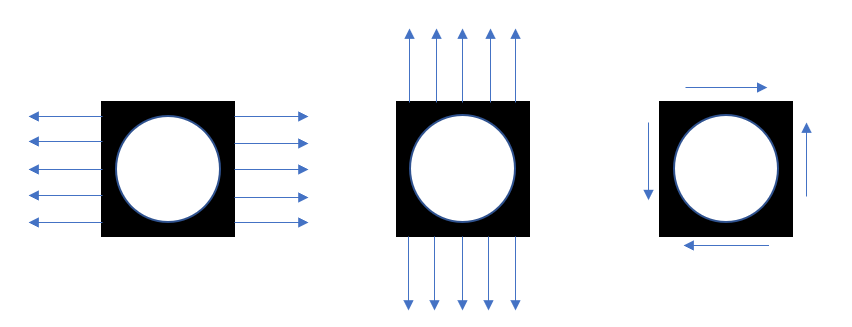}%
		\caption{Extracting the elasticity tensor.}
		\label{fig:Homogenization}
	\end{center}
\end{figure}

Given a unit cell (microstructure) in 2D, a forward problem is to find its homogenized elasticity $3 \times 3$ tensor  $\boldsymbol{C}^H$. The theory of homogenization is well developed, for example, see \cite{yang2020determination, hassani1998review}. A typical numerical strategy \cite{andreassen2014determine} for computing  $\boldsymbol{C}^H$ is to impose three independent periodic boundary conditions (in 2D), and solve the resulting  finite element problems; see Figure \ref{fig:Homogenization}.  The stresses and strains from the three problems are then used to compute  $\boldsymbol{C}^H$ as described in \cite{andreassen2014determine}.

Next, various quantities of interest, namely,  bulk modulus ($K$), shear modulus ($G$) and Poisson ratio  ($\nu$) can be extracted from $\boldsymbol{C}^H$ as follows:

\begin{subequations}
	\begin{align}
	 K = (C^H_{1,1} + C^H_{2,2} +C^H_{1,2} +C^H_{2,1})/4
    \label{eq:BulkModulus}\\
     G = C^H_{3,3} 
    \label{eq:ShearModulus}\\
   \nu = (C^H_{2,1} + C^H_{1,2})/(C^H_{1,1} +C^H_{2,2}) 
    \label{eq:PoissonRatio}
	\end{align}
\end{subequations}
\subsection{Microstructural Optimization Methods}
\label{subsec:litreview}

The inverse problem is to arrive at an optimal topology that  maximizes or minimizes one of these quantities of interest. There are several methods available today for optimizing microstructures; see, for example, \cite{osanov2016topology, gao2020topology, vogiatzis2017topology, SureshMicrostructural, kollmann2020deep, guo2020semi}.

In the  popular density-based methods,  one defines a pseudo-density  $\rho_e\in(0,1]$ over the underlying finite element mesh. Then the microstructural optimization problem, of say, maximizing the bulk modulus, subject to a mass constraint (we prefer here a mass constraint over a volume constraint since this generalizes more easily to multiple materials), may be posed as follows:

\begin{subequations}
	\begin{align}
	& \underset{\boldsymbol{\rho}}{\text{minimize}}
	& &-K(\boldsymbol{\rho})\label{eqnObj_microstr}\\
	& \text{subject to}
	& & \mathbf{K}(\boldsymbol{\rho})\boldsymbol{u}^i = \boldsymbol{f}^i, i = 1,2,3\label{eqn:GoverningEqn_microstr}\\
	& & & \sum_e \rho_e v_e \lambda_e = \hat{m}\label{eqn:masscons_microstr}\\
	& & & 0 < \rho_e \le 1 \label{eqn:densityConstraint}
	\end{align}
\end{subequations}
where  $\mathbf{K}$ is the  stiffness matrix, $\boldsymbol{u}^i$ and $\boldsymbol{f}^i$ are the displacement vector and the external force vector for the three problems in Figure ~\ref{fig:Homogenization}, $v_e$ is the volume of the finite element,  $\lambda_e$ is the physical density of the base material, and $\hat{m}$ is the mass constraint. In addition, the solid isotropic material with penalization (SIMP) penalization model is employed to  link the density variables to the base material  
\begin{equation}
    E(\rho_e) = E_{min} + E \rho_e^p
    \label{eq:SIMPModel}
\end{equation}
The field can now be optimized using, for example, optimality criteria \cite{Bendsoe2003} or MMA \cite{svanberg1987MMA}, resulting in the desired microstructural topology; see \cite{xia2015design}, for example. Similar problems can be posed, for example, to minimize the Poisson ratio, subject to a mass constraint. 

\section{Proposed Framework}
\label{sec:methodology}

\subsection{Generalized Microstructural Problems}
\label{subsec:generalizedProblems}

As stated earlier, in many applications, the  mass constraint $\hat{m}$ (see Equation \ref{eqn:masscons_microstr}) is not known a priori.  Instead, it may be more advantageous to impose constraints on physical quantities. In this paper, we consider such generalized microstructural optimization problems of the form:

\begin{subequations}
	\begin{align}
	& \underset{\boldsymbol{\rho}}{\text{minimize}}
	& &\phi( \boldsymbol{C}^H (\boldsymbol{\rho}))\label{eqnObj_microstr}\\
	& \text{subject to}
	& & \boldsymbol{K}(\boldsymbol{\rho})\boldsymbol{u}^i = \boldsymbol{f}^i, i = 1,2,3\label{eqn:GoverningEqn_microstr}\\
	& & & h_j(\boldsymbol{\rho}) = 0, j = 1,2,... \label{eqn:equalityConstraint_microstr}\\
	& & & 0 < \boldsymbol{\rho} \le 1 \label{eqn:densityConstraint_microstr}
	\end{align}
\end{subequations}
where the objective $\phi( \boldsymbol{C}^H)$ represents one of the following: the bulk modulus ($K$), shear modulus ($G$), Poisson ratio  ($\nu$), or mass ($m$), and $h_j$ are equality constraints involving these quantities. 

For example, the problem of maximizing the bulk modulus subject to a Poisson ratio constraint may be posed as:
\begin{subequations}
	\begin{align}
	& \underset{\boldsymbol{\rho}}{\text{minimize}}
	& &-K(\boldsymbol{\rho})\label{eqnObj_microstr}\\
	& \text{subject to}
	& & \boldsymbol{K}(\boldsymbol{\rho})\boldsymbol{u}^i = \boldsymbol{f}^i, i = 1,2,3\label{eqn:GoverningEqn_microstr}\\
	& & & \nu(\boldsymbol{\rho})/\hat{\nu} -1 = 0 \label{eqn:volcons_microstr}\\
	& & & 0 < \boldsymbol{\rho} \le 1 \label{eqn:densityConstraint_microstr}
	\end{align}
\end{subequations}
where $\hat{\nu}$ is the desired Poisson ratio. Similarly, the problem of minimizing the mass  subject to constraints on the bulk modulus and Poisson ratio  may be posed as:
\begin{subequations}
	\begin{align}
	& \underset{\boldsymbol{\rho}}{\text{minimize}}
	& &m(\boldsymbol{\rho})\label{eqnObj_microstr}\\
	& \text{subject to}
	& & \boldsymbol{K}(\boldsymbol{\rho})\boldsymbol{u}^i = \boldsymbol{f}^i, i = 1,2,3\label{eqn:GoverningEqn_microstr}\\
	& & & K(\boldsymbol{\rho})/\hat{K} -1 = 0 \label{eqn:GoverningEqn_microstr}\\
	& & & \nu(\boldsymbol{\rho})/\hat{\nu} -1 = 0 \label{eqn:volcons_microstr}\\
	& & & 0 < \boldsymbol{\rho} \le 1 \label{eqn:densityConstraint_microstr}
	\end{align}
\end{subequations}
where $\hat{K}$ is the desired bulk modulus, and $\hat{\nu}$ is the desired Poisson ratio. In the remainder of this paper, we will consider solving such problems.

\subsection{Representing Density using Neural Networks}
\label{sec:method_DesignRepresentationNN}

One of the challenges in solving such generalized problems is computing the sensitivities of the objective and constraints. Manual derivation  can be cumbersome and error-prone, especially when the problem is recast within the context of augmented Lagrangian formulation (see Section \ref{subsec:augLag}). We, therefore, propose here a neural-network (NN) framework that supports automatic sensitivity computation for gradient-based optimization \cite{ChandrasekharAuTO2021}. The framework not only eliminates the burden of manual sensitivity calculations, it offers other computational advantages as discussed in the remainder of the paper. 

\begin{figure}[H]
	\begin{center}
    \includegraphics[scale=0.5]{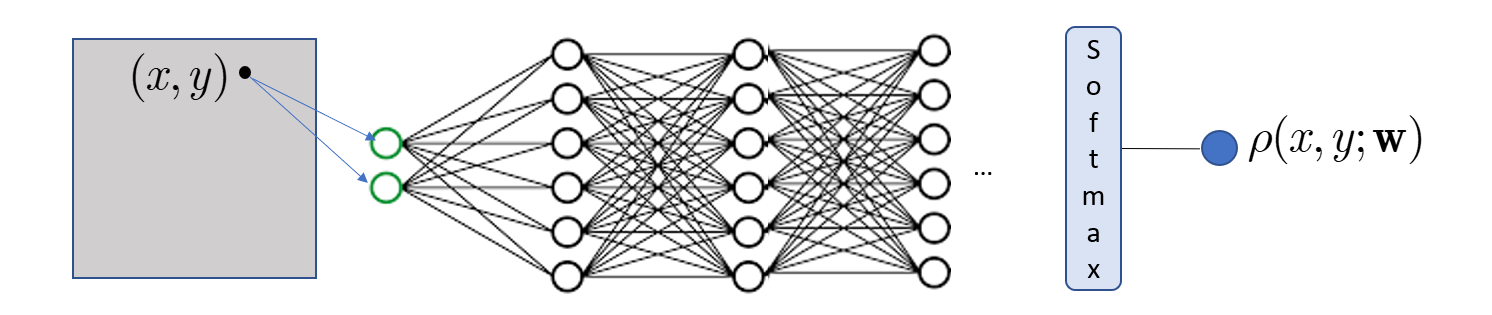}%
		\caption{Neural network architecture for representing the density field.}
		\label{fig:NNarchitecture}
	\end{center}
\end{figure}

In particular, we employ a simple fully-connected feed-forward neural network \cite{bishop2006pattern}. The input to the network are points ($x$, $y$) within the domain; the NN has a series of hidden layers associated with activation functions  \cite{Goodfellow2016_deepLearning}, \cite{lu2019dyingReLU}. The output is the density  $\rho$ at that point; see Figure \ref{fig:NNarchitecture}. Observe that the final layer is  a SoftMax activation function  \cite{bishop2006pattern} that ensures that the  density lies between 0 and 1. The density  will depend on the weights and bias, i.e., they serve as the design variables $\boldsymbol{w}$. 

Thus, we can now repose the generalized problem as:

\begin{subequations}
	\begin{align}
	& \underset{\boldsymbol{w}}{\text{minimize}}
	& &\phi( \boldsymbol{C}^H (\boldsymbol{w}))\label{eq:eqnObj_microstr}\\
	& \text{subject to}
	& & \boldsymbol{K}(\boldsymbol{w})\boldsymbol{u}^i = \boldsymbol{f}^i, i = 1,2,3\label{eqn:GoverningEqn_microstr}\\
	& & & h_j(\boldsymbol{w}) = 0, j = 1,2,...n \label{eq:volcons_microstr}
	\end{align}
	\label{eq:general_NN_eqn}
\end{subequations}
Observe that an explicit (bound) constraint on the density field is not needed since it is automatically satisfied by the Softmax function.

\begin{figure}[H]
	\begin{center}
    \includegraphics[scale=0.55]{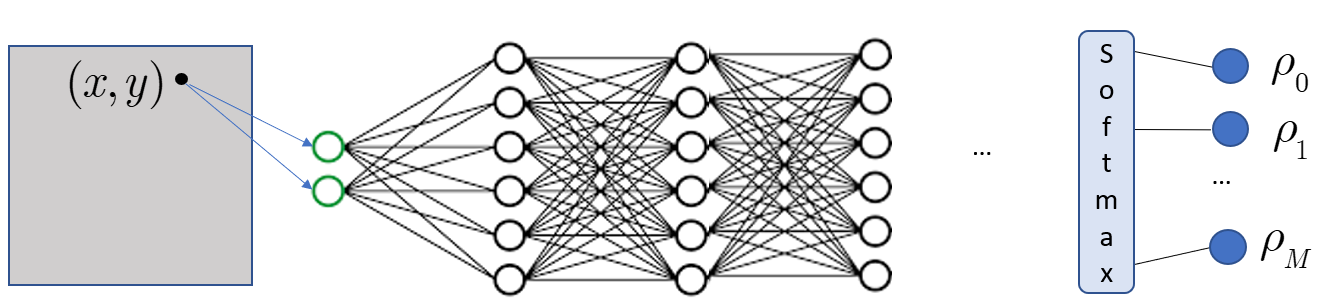}%
		\caption{Neural network architecture for representing multiple materials.}
		\label{fig:NNarchitectureMultiple}
	\end{center}
\end{figure}

To generalize the above framework  to multiple materials, the output of the NN is increased to $(M+1)$ variables, where $M$ is the number of non-void materials; see   Figure \ref{fig:NNarchitectureMultiple}. No other change is needed in the framework, i.e., one can use exactly the same number of design variables $\boldsymbol{w}$. Further, due to the nature of Softmax function, the partition of unity condition:
\begin{equation}
    \sum\limits_{m=0}^M  \rho_m = 1
    \label{eq:PartitionOfUnity}
\end{equation}
is  automatically satisfied, i.e., the sum of all densities is guaranteed to be unity. Finally, the SIMP material model in Equation  \ref{eq:SIMPModel} is generalized to multiple materials  as follows:
\begin{equation}
    E(\rho) = E_{min} +\sum\limits_{m=0}^M  E_m\rho_m^p
    \label{eq:SIMPMultipleMaterialModel}
\end{equation}

\subsection{Augmented Lagrangian}
\label{subsec:augLag}

\begin{figure}[H]
	\begin{center}
    \includegraphics[scale=0.65]{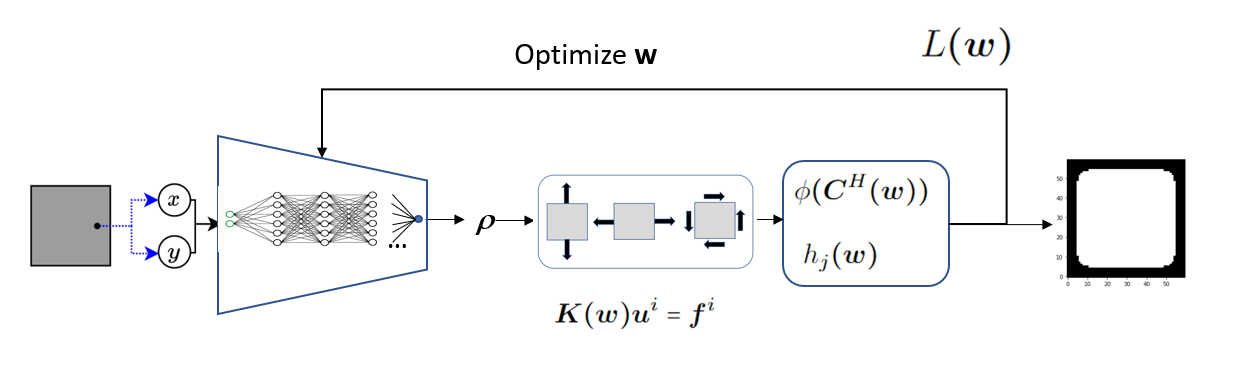}%
		\caption{The optimization framework.}
		\label{fig:optimizationFramework}
	\end{center}
\end{figure}

To solve both the single and multi-material  problems, we will use  the augmented Lagrangian method \cite{nocedal2006numericalOptimization}, i.e., let
\begin{equation}
L(\boldsymbol{w}) = \phi + \sum\limits_{j=1,2 ...}^{n} {\alpha_j} h_j^2 + \sum\limits_{j=1,2 ...}^{n} {\mu_j} h_j
\label{eq:lossEqn}
\end{equation}
where the penalty parameters $\alpha_j$ and Lagrange multipliers $\mu_j$ are updated through iterations. First, starting with a small value for $\alpha_j$ and a zero value for $\mu_j$  the augmented Lagrangian is minimized using the popular L-BFGS method \cite{nocedal2006numericalOptimization}. Then,  the coefficients $\alpha$ and $\mu$ are updated as follows (see Section \ref{sec:experiments}):
\begin{subequations}
	\begin{align}
	 {\alpha_j}^{(k+1)} =  {\alpha_j}^{(k)}  + \Delta \alpha
    \label{eq:augLagAlphaUpdate}\\
     {\mu_j}^{(k+1)}  = {\mu_j}^{(k)}  + 2{\alpha_j}^{(k+1)}  h_j^{(k)} 
    \label{eq:augLagMuUpdate}
	\end{align}
\end{subequations}
The optimization is repeated until convergence. For termination, we consider two quantities
\begin{equation}
\epsilon_{\phi} = \mid (\phi^{(k+1)} - \phi^{(k)})/\phi^{(k+1)}\mid
  \label{eq:phiError}
\end{equation}
and
\begin{equation}
\epsilon_{h} = \sum\limits_{j=1,2 ...}^{n}\mid h_j \mid
\label{eq:gError}
\end{equation}
The algorithm terminates when both quantities are less than a prescribed value (see Section \ref{sec:experiments} for details). The overall framework is illustrated in Figure \ref{fig:optimizationFramework}.

\subsection{Algorithm}
\label{subsec:algorithm}

The algorithm for the proposed framework is described in \cref{alg:MicroTOuNN}. First, the target design objective $\phi$  and the set of equality constraints $h_j$ are specified along with an initial topology. Note that defining an initial topology is equivalent to initializing the neural network. Next, the domain is discretized into finite elements and the element centers serve as input to the neural network, which returns the elemental densities $\bm{\rho}$ as discussed in \cref{sec:method_DesignRepresentationNN}.  Homogenization is performed which predicts the homogenized structural response from the elemental densities as per \cref{subsec:homogenization}. Based on the results from homogenization, the objective $\phi$ \cref{eq:eqnObj_microstr} and the constraints $h_j$ \cref{eq:volcons_microstr} are evaluated.  Next, the loss function $L$ is computed using \cref{eq:lossEqn}, and the gradient of the loss function with respect to the neural network weights ($\nabla L$) is computed through automatic differentiation. The NN is then trained using PyTorch's L-BFGS optimizer as described in \cref{subsec:augLag}, which requires the update of Lagrangian parameters $\alpha,\mu$. The training procedure continues until the convergence criteria  are met.

\begin{algorithm}[H]
	\caption{Microstructural TO}
	\label{alg:MicroTOuNN}
	\begin{algorithmic}[1]
		\Procedure{MicroTOuNN}{$\Omega_0$, $\phi$, \ldots} \Comment{\parbox[t]{.3\linewidth}{Inputs}}
		
		\State $\Omega^0 \rightarrow \Omega^0_h$ \Comment{Domain discretization} \label{alg:domainDiscretize}
		
		\State $\bm{x} = \{x_e,y_e\}_{e \in \Omega^0_h} $ \Comment{elem centers; NN input} \label{alg:elemCenterComp}
		
		\State  k = 0; $\alpha_i = \alpha_0$; $\mu_i = 0 \; \forall i$ \Comment{Initialization (\cref{subsec:augLag})}
		\label{alg:Initialization }
		
		\Repeat \Comment{Optimization (Training)}
		
		\State $NN(\bm{x} ; \bm{w}) \rightarrow \bm{\rho}$ \Comment{Fwd prop NN} \label{alg:fwdPropNN}

        \State $ \bm{\rho} \rightarrow \boldsymbol{C}^H$ \Comment{Homogenization (\cref{subsec:homogenization})} \label{alg:homogenization}
  
        \State $ \boldsymbol{C}^H \rightarrow \phi $\Comment{Compute design objective \cref{eq:eqnObj_microstr}} \label{alg:objective}
          
  		\State $\{\boldsymbol{C}^H, \bm{\rho} \} \rightarrow h_j $ \Comment{Compute  constraints \cref{eq:volcons_microstr}} \label{alg:massConstraint}
  		
		\State $\{\phi, h, \alpha_i, \mu_i\} \rightarrow L$ \Comment{Loss from \Cref{eq:lossEqn}} \label{alg:lossCompute}
		
		\State $AD(L \leftarrow \bm{w}) \rightarrow \nabla L $ \Comment{Auto diff for sensitivity} \label{alg:autoDiff}
			 
		\State $\bm{w}  +  \Delta \bm{w} (\nabla L ) \rightarrow \bm{w} $ \Comment{L-BFGS  step} \label{alg:LBFGSStep}
		
		\State $\text{k}++$
		
		\State Update $\alpha_i, \mu_i$ \Comment {Aug-Lag terms \cref{eq:augLagAlphaUpdate}, \cref{eq:augLagMuUpdate}} \label{alg:OptPenaltyUpdate}

		\Until{ $ \Delta \phi < \hat{\epsilon}_\phi$ and $ \Delta h < \hat{\epsilon}_h $  and $k < k_{max}$} \Comment{convg. criteria}
		
	    \State	\Return $\bm{w}$

		\EndProcedure
	\end{algorithmic}
\end{algorithm}

\section{Numerical examples}
\label{sec:experiments}

In this section, we conduct several numerical experiments to illustrate the proposed algorithm. The implementation is in Python, within the PyTorch environment.  All experiments were conducted on an Intel i9-11900K @ 3.50GHz GHz, equipped with 32 GB of RAM. The default parameters are listed in Table \ref{table:defaultParameters}. We use the popular SIMP continuation scheme \cite{Rojas2015Continuation} for the material model. For the default NN configuration of 5 layers and 30 neurons per layer, the number of design variables $\bm{w}$ (weight and bias) is 3872.

\begin{figure}[H]
	\begin{center}
    \includegraphics[scale=0.5]{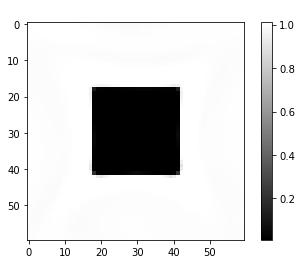}%
		\caption{Default initial topology.}
		\label{fig:initialTopology}
	\end{center}
\end{figure}

\begin{table} [h!]
	\begin{center}
		\begin{tabular}{  r | l   }
			Parameter & Description and default value \\ \hline
			$E$, $\nu$ & For base material: $E=1$, $\nu=0.3$; for void: $E=0.001$, $\nu=0.3$  \\
		    $\lambda$(mass density) & For base material: $\lambda = 1$;  for void: $\lambda = 0.001$  \\
			NN &  Neural network: 5 layers and 30 neurons per layer, with Swish functions  \\
			$\alpha$, $\mu$ & Lagrangian parameters: $\alpha_0 = 1,  \Delta \alpha = 5$ and $\mu_0 = 0$\\
			$p$ & SIMP continuation parameters: $p_0 = 2$, $\Delta p = 0.5$, $ p_{max} = 10$ \\
			$n_x$,$n_y$ & Mesh elements: $n_x = 60$, $n_y = 60$ \\
			$\hat{\epsilon}_\phi$, $\hat{\epsilon}_h$ & Convergence criteria for objective and constraints: $\hat{\epsilon}_\phi$ = $\hat{\epsilon}_h$ = 0.025  \\

		\end{tabular}
	\end{center}
	\caption{Default simulation parameters.}
	\label{table:defaultParameters}
\end{table}
The default initial topology is a square block as illustrated in Figure \ref{fig:initialTopology}.

 Through the experiments, we investigate the following:
\begin{enumerate}
	\item $\textit{Validation}$: First, we consider classic mass constrained optimization of bulk modulus, shear modulus, and Poisson ratio, and compare some of the computed values against well-established theoretical results.
	
	\item $\textit{Convergence}$: The typical convergence of the algorithm is then illustrated.
		
	\item $\textit{Initial Topology}$: Next, we consider initial topologies different from the one in Figure \ref{fig:initialTopology} and optimize for maximum bulk modulus.
	
	\item $\textit{Impact of NN Configuration}$: We then vary the NN size and consider its impact on Poisson ratio minimization.
	
	\item $\textit{Impact of Mesh Size}$: We repeat the above experiment but vary the mesh size instead of the NN size.
	
	\item $\textit{Generalized Problems}$: We then consider several generalized microstructural problems.
    
    \item $\textit{Multi-material design}$: Finally, we present results for multiple materials.

\end{enumerate}

\subsection{Validation}
\label{sec:expts_validation}

\begin{figure}[H]
\begin{subfigure}[b]{.31\textwidth}
  	\begin{center}
    \includegraphics[scale=0.3]{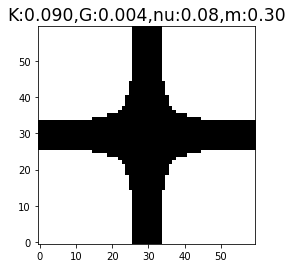}%
	\end{center}
	\caption{Bulk modulus.}
	\label{fig:maxBulkModulus}
		\end{subfigure}
	\begin{subfigure}[b]{.31\textwidth}
  	\begin{center}
    \includegraphics[scale=0.3]{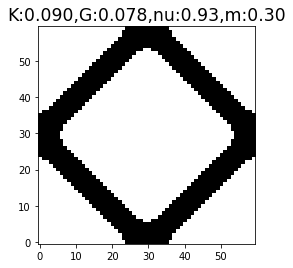}%
		\caption {Shear modulus.}
		\label{fig:maxShearModulus}
	\end{center}
	\end{subfigure}
		\begin{subfigure}[b]{.31\textwidth}
  	\begin{center}
    \includegraphics[scale=0.3]{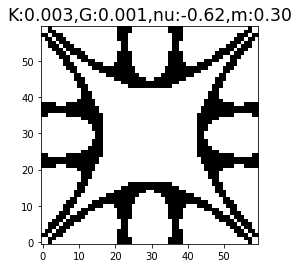}%
		\caption{Poisson ratio.}
		\label{fig:minPoissonRatio}
	\end{center}
	\end{subfigure}
	\caption{Classic microstructural optimization problems.}
\end{figure}

First, we consider the classic problem of  bulk modulus maximization subject to a mass-fraction constraint of 0.3. The resulting topology, computed in 47 seconds, is illustrated in Figure \ref{fig:maxBulkModulus}. Similarly,  Figure \ref{fig:maxShearModulus} illustrates a  microstructure with maximal shear modulus for the same mass constraint, computed in 53 seconds. Finally,  Figure \ref{fig:minPoissonRatio} illustrates the microstructure when the Poisson ratio is minimized for the same mass constraint, computed in 55 seconds,

Table \ref{table:maxK} compares the computed bulk modulus for various mass fractions, against the Hashin-Shtrikman upper bound.
\begin{table}[ht]
	\centering
    \begin{tabular}{||l|l|l||}
			\hline
			$\hat{m}$ & $K^*$ &  $K_{HS}$  \\ \hline
			0.1 & 0.020 & 0.0282   \\ \hline
			0.3 & 0.092 & 0.099  \\ \hline
			0.5 & 0.18 & 0.20  \\ \hline
			0.7 & 0.34 & 0.35  \\  \hline
    \end{tabular}
    \caption{Computed bulk modulus vs. HS upper bound.  }
    \label{table:maxK}
\end{table}

\subsection{Convergence}

\begin{figure}[H]
  \begin{subfigure}[b]{.45\textwidth}
  	\begin{center}
    \includegraphics[scale=0.4]{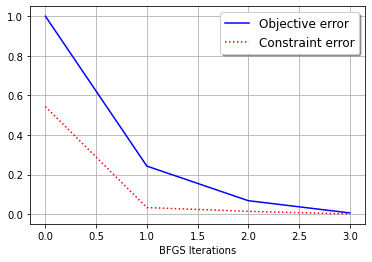}%
		\caption{Bulk modulus.}
		\label{fig:KError}
	\end{center}
	\end{subfigure}
	\begin{subfigure}[b]{.45\textwidth}
  	\begin{center}
    \includegraphics[scale=0.4]{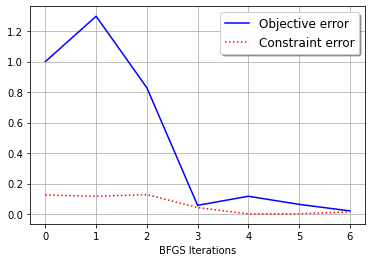}%
		\caption{Poisson ratio.}
		\label{fig:nuError}
	\end{center}
	\end{subfigure}
	\caption{Objective and constraint errors after every L-BFGS iteration.}
	\label{fig:ErrorPlots}
\end{figure}

Recall that the error in the objective $\epsilon _ \phi$ and error in the constraints $\epsilon _ g$, defined in Equation \ref{eq:phiError} and Equation\ref{eq:gError} respectively, are computed at the end of every L-BFGS iteration. The algorithm terminates when  $	\epsilon _ \phi  < \hat{\epsilon}_\phi$ \emph {and} $ \epsilon _ g < \hat{\epsilon}_g$ specified in Table \ref{table:defaultParameters}. Figure \ref{fig:KError} illustrates the objective error $\epsilon _ \phi $ and the constraint error $\epsilon _ g$  for bulk modulus maximization, while Figure \ref{fig:nuError} illustrates the two errors for Poisson ratio minimization. 

\begin{figure}[H]
  \begin{subfigure}[b]{.45\textwidth}
  	\begin{center}
    \includegraphics[scale=0.425]{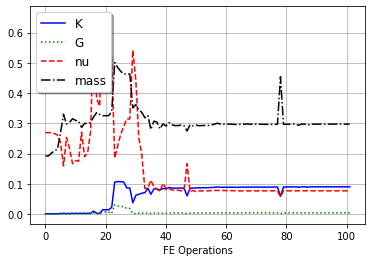}%
		\caption{Bulk modulus.}
		\label{fig:KConvergence}
	\end{center}
	\end{subfigure}
	\begin{subfigure}[b]{.45\textwidth}
  	\begin{center}
    \includegraphics[scale=0.425]{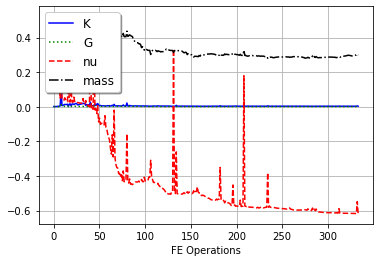}%
		\caption{Poisson ratio.}
		\label{fig:nuConvergence}
	\end{center}
	\end{subfigure}
	\caption{Various homogenized quantities after every FE operation.}
	\label{fig:ErrorPlots}
\end{figure}

In the experiments, we observed that 3 to 20 L-BFGS iterations are sufficient. Further, each L-BFGS iteration typically involves 2 to 10 inner iterations. The homogenized properties are recorded at the end of every inner iteration. Figure \ref{fig:KConvergence} illustrates these quantities for bulk modulus maximization, and Figure \ref{fig:nuConvergence} illustrates these for Poisson ratio minimization. The spikes observed in the two figures correspond to the start of a new L-BFGS iteration when the penalty parameters are updated.

\subsection{Impact of Initial Topology}
\label{sec:initialTopology}
Next, we  consider initial designs different from Figure \ref{fig:initialTopology} and study their impact on the optimal topology, with all other parameters are kept constant at default values. Specifically, we revisit the problem of maximizing the bulk modulus, subject to a mass-fraction constraint of 0.3. 

\begin{figure}[H]
\begin{center}
  \begin{subfigure}[b]{.3\textwidth}
  	\begin{center}
    \includegraphics[scale=0.3]{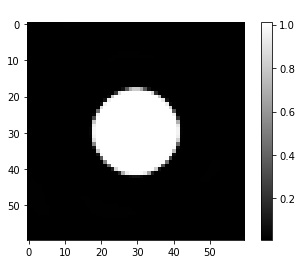}%
		\caption{Initial topology.}
		\label{fig:circularHole}
	\end{center}
	\end{subfigure}
    \begin{subfigure}[b]{.3\textwidth}
  	\begin{center}
     \includegraphics[scale=0.3]{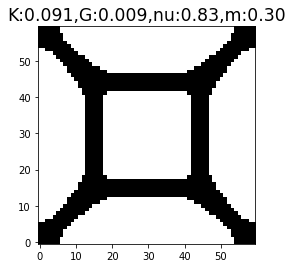}%
		\caption {Final topology.}
		\label{fig:circularHoleMaxK}
	\end{center}
	\end{subfigure}
	\caption{Maximizing bulk modulus with mass constraint.}
	\end{center}
\end{figure}

For the initial topology of a circular hole in Figure \ref{fig:circularHole}, the final topology is illustrated in  Figure \ref{fig:circularHoleMaxK}.
Similarly, with the initial topology of four circular holes in Figure \ref{fig:circularHoles}, the final topology, for the same problem is illustrated in  Figure \ref{fig:circularHolesMaxK}, with $K^*=0.090$. 

\begin{figure}[H]
\begin{center}
  \begin{subfigure}[b]{.3\textwidth}
  	\begin{center}
    \includegraphics[scale=0.3]{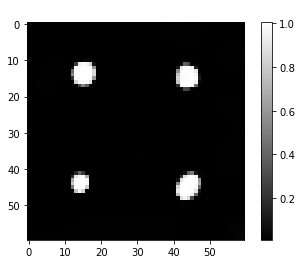}%
		\caption{Initial topology.}
		\label{fig:circularHoles}
	\end{center}
	\end{subfigure}
    \begin{subfigure}[b]{.3\textwidth}
  	\begin{center}
     \includegraphics[scale=0.3]{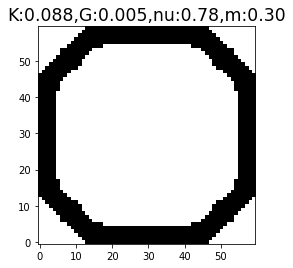}%
		\caption {Final topology.}
		\label{fig:circularHolesMaxK}
	\end{center}
	\end{subfigure}
	\caption{Maximizing bulk modulus with mass constraint.}
	\end{center}
\end{figure}

Although the designs are different, the final bulk moduli are comparable. Thus, one can potentially discover different designs by changing the initial design.

\subsection{Impact of NN Size}
\label{sec:NNSize}
\begin{figure}[H]
\begin{center}
  \begin{subfigure}[b]{.3\textwidth}
  	\begin{center}
    \includegraphics[scale=0.3]{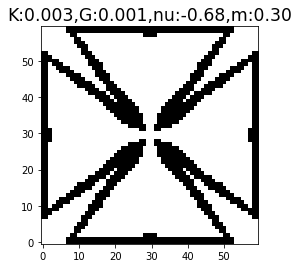}%
		\caption{NN: $5 \times 10$.}
		\label{fig:circularHoles}
	\end{center}
	\end{subfigure}
    \begin{subfigure}[b]{.3\textwidth}
  	\begin{center}
    \includegraphics[scale=0.3]{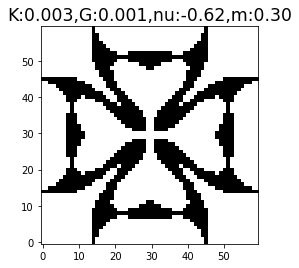}%
		\caption {NN: $5 \times 20$.}
		\label{fig:circularHolesMaxK}
	\end{center}
	\end{subfigure}
	\begin{subfigure}[b]{.3\textwidth}
  	\begin{center}
    \includegraphics[scale=0.3]{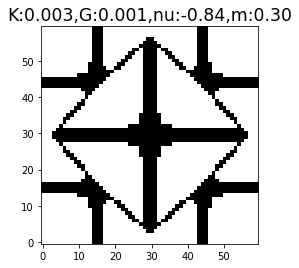}%
		\caption {NN: $8 \times 40$.}
		\label{fig:circularHolesMaxK}
	\end{center}
	\end{subfigure}
	\caption{Minimizing Poisson ratio with mass constraint using different NNs.}
	\label{fig:ImpactOfNN}
	\end{center}
\end{figure}

We now vary the neural-network size and minimize the Poisson ratio, subject to a mass-fraction constraint of 0.3 (with default parameters).  Figure \ref{fig:ImpactOfNN} illustrates various designs obtained for different NN configurations. The final Poisson ratio $\nu^*$ for the three cases are -0.68, -0.62 and -0.84 respectively. The number of design variables  are 492, 1782 and 11,682 respectively. We observe that as the number of design variables is increased, the algorithm tends to generate more complex designs. However, we did not observe any pattern between the NN size and the final objective achieved.

\subsection{Impact of Mesh Size}
\label{sec:MeshSize}

\begin{figure}[H]
\begin{center}
  \begin{subfigure}[b]{.3\textwidth}
  	\begin{center}
    \includegraphics[scale=0.3]{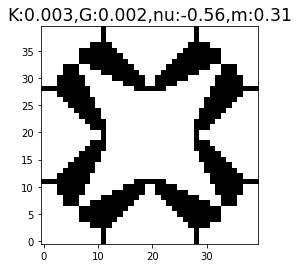}%
		\caption{Mesh: $40 \times 40$.}
		\label{fig:circularHoles}
	\end{center}
	\end{subfigure}
    \begin{subfigure}[b]{.3\textwidth}
  	\begin{center}
    \includegraphics[scale=0.3]{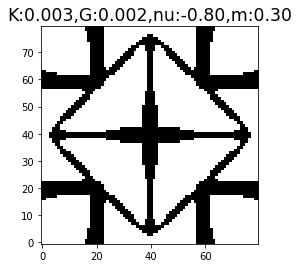}%
		\caption {Mesh: $80 \times 80$.}
		\label{fig:circularHolesMaxK}
	\end{center}
	\end{subfigure}
	\begin{subfigure}[b]{.3\textwidth}
  	\begin{center}
    \includegraphics[scale=0.3]{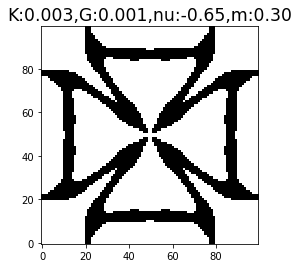}%
		\caption {Mesh: $100 \times 100$.}
		\label{fig:circularHolesMaxK}
	\end{center}
	\end{subfigure}
	\caption{Minimizing Poisson ratio for different mesh sizes.}
	\label{fig:ImpactOfMeshSize}
	\end{center}
\end{figure}

Next, we vary the mesh size and minimize the Poisson ratio, subject to a mass constraint of 0.3 (with all other parameters at default).  Figure \ref{fig:ImpactOfMeshSize} illustrates  the designs obtained for different mesh sizes. The final Poisson ratio $\nu^*$ for the three cases are -0.56, -0.80 and -0.65 respectively. Once again, we did not observe any pattern between the mesh size and the final objective achieved.

\subsection{Generalized Problems}
\label{sec:bulkModulusConstraint}
\begin{figure}[H]
\begin{center}
  \begin{subfigure}[b]{.3\textwidth}
  	\begin{center}
    \includegraphics[scale=0.5]{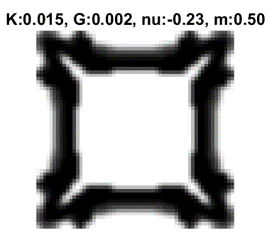}%
		\caption{Mass constrained.}
		\label{fig:classic}
	\end{center}
	\end{subfigure}
    \begin{subfigure}[b]{.3\textwidth}
  	\begin{center}
    \includegraphics[scale=0.45]{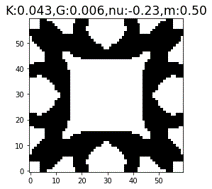}%
		\caption {Maximizing $K$.}
		\label{fig:maxKGeneralized}
	\end{center}
	\end{subfigure}
	\begin{subfigure}[b]{.3\textwidth}
  	\begin{center}
    \includegraphics[scale=0.45]{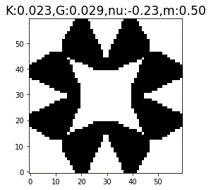}%
		\caption {Maximizing $G$.}
		\label{fig:maxGGeneralized}
	\end{center}
	\end{subfigure}
	\caption{Classic versus proposed framework.}
	\label{fig:ClassicVersusProposed}
	\end{center}
\end{figure}

To illustrate the benefits of the proposed framework, we first used the MATLAB code published in \cite{xia2015design} to minimize the Poisson ratio to a mass constraint of $\hat{m} =0.5$. The algorithm did not terminate; the final topology, after a maximum allowable 450 FE operations (15 seconds), is illustrated in Figure \ref{fig:classic}; it exhibits the following characteristics $\nu = -0.23$, $K =0.015$ and $G = 0.002$. Observe that the microstructure is weak in bulk and shear. 

In the proposed framework, the bulk modulus was maximized subject to $\nu = -0.23$ and $\hat{m} = 0.5$, i.e., the constraints are consistent with the results obtained above. The resulting topology, after 420 FE operations (52 seconds), illustrated in Figure \ref{fig:maxKGeneralized} exhibits the following characteristics $m =0.5$, $\nu = -0.23$, $K =0.043$ and $G = 0.006$. Thus the proposed framework increases the bulk and shear moduli by a factor of $3X$. For approximately the same number of finite element operations, the proposed framework is slower due to the overhead of automatic differentiation.

Next, the shear modulus was maximized subject to $\nu = -0.23$ and $\hat{m} = 0.5$. The resulting topology after 560 FE operations (91 seconds),  illustrated in Figure \ref{fig:maxGGeneralized} exhibits the following characteristics $m =0.5$, $\nu = -0.23$, $K =0.023$ and $G = 0.029$, i.e., the bulk and shear moduli increased by a factor of $10X$.

\begin{figure}[H]
\begin{center}
  \begin{subfigure}[b]{.3\textwidth}
  	\begin{center}
    \includegraphics[scale=0.3]{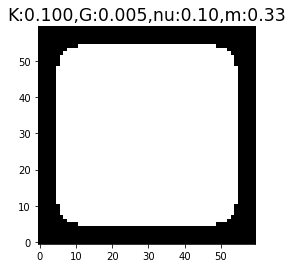}
		\caption{$\hat{K} = 0.1$, $\hat{\nu} = 0.1$.}
		\label{fig:circularHoles}
	\end{center}
	\end{subfigure}
    \begin{subfigure}[b]{.3\textwidth}
  	\begin{center}
    \includegraphics[scale=0.3]{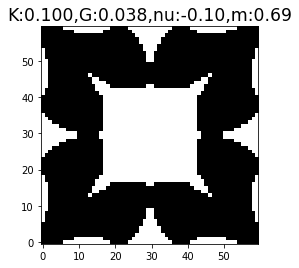}
		\caption{$\hat{K} = 0.1$, $\hat{\nu} = -0.1$.}
		\label{fig:circularHolesMaxK}
	\end{center}
	\end{subfigure}
	\begin{subfigure}[b]{.3\textwidth}
  	\begin{center}
    \includegraphics[scale=0.3]{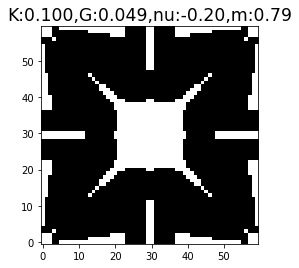}
		\caption{$\hat{K} = 0.1$, $\hat{\nu} = -0.2$.}
		\label{fig:circularHolesMaxK}
	\end{center}
	\end{subfigure}
	\caption{Minimizing mass with bulk modulus and Poisson ratio constraints.}
	\label{fig:MultipleConstraints}
	\end{center}
\end{figure}

\begin{figure}[H]
\begin{center}
  \begin{subfigure}[b]{.3\textwidth}
  	\begin{center}
    \includegraphics[scale=0.3]{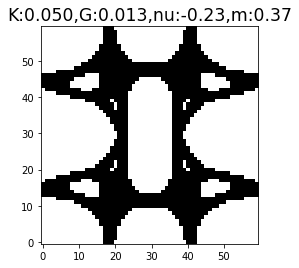}
		\caption{m = 0.37}
		\label{fig:target1}
	\end{center}
	\end{subfigure}
    \begin{subfigure}[b]{.3\textwidth}
  	\begin{center}
    \includegraphics[scale=0.3]{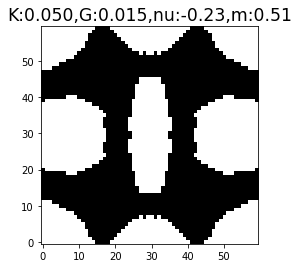}
	\caption{m = 0.51}
		\label{fig:target1}
	\end{center}
	\end{subfigure}
	\begin{subfigure}[b]{.3\textwidth}
  	\begin{center}
    \includegraphics[scale=0.55]{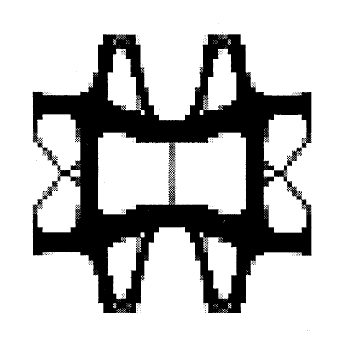}
		\caption{m = 0.37 \cite{yin2001optimality}}
		\label{fig:reference}
	\end{center}
	\end{subfigure}
	\caption{Designs obtained for a targeted $\boldsymbol{\hat{C}}^H$.}
	\label{fig:targetCSolutions}
	\end{center}
\end{figure}

One can also impose multiple physical constraints within the proposed framework. Consider the minimization of the mass, subject to bulk modulus and Poisson ratio constraints. Figure \ref{fig:MultipleConstraints} illustrates three  designs obtained for three different scenarios. The corresponding mass fractions are 0.33, 0.69, and 0.79. 
One can also target a specific elasticity matrix $\boldsymbol{\hat{C}}^H$. We consider  a specific example from \cite{yin2001optimality}:
\begin{equation}
\boldsymbol{\hat{C}}^H = \begin{bmatrix}
0.13 & -0.03 & 0\\
-0.03 &  0.13 & 0\\
0 & 0 & 0.015
\end{bmatrix}
\label{eq:targetC}
\end{equation}
Note that there may exist multiple solutions (with significantly different masses) with the same $\boldsymbol{C}^H$ \cite{vogiatzis2017topology}. Using our framework by imposing the target $\boldsymbol{C}^H$ and a target mass as constraints, one can obtain different designs; these are illustrated in Figure \ref{fig:targetCSolutions}a and \ref{fig:targetCSolutions}b, with mass target of 0.37 and 0.51, respectively. A solution, with a mass of 0.37 was reported  in \cite{yin2001optimality}; see Figure \ref{fig:targetCSolutions}c.

\subsection{Sampling at Higher Resolution}
\label{sec:multiMaterial}

\begin{figure}[H]
\begin{center}
	\begin{subfigure}[b]{.4\textwidth}
  	\begin{center}
    \includegraphics[scale=0.4]{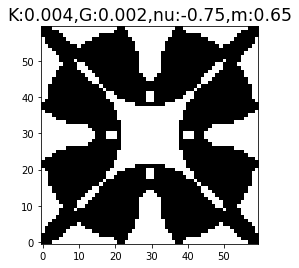}%
		\caption {Raw output.}
	\end{center}
	\end{subfigure}
		\begin{subfigure}[b]{.4\textwidth}
  	\begin{center}
    \includegraphics[scale=0.4]{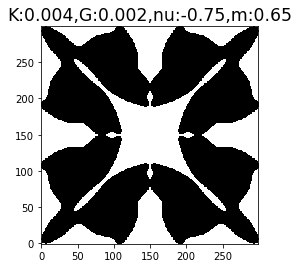}%
		\caption {Sampling at higher resolution.}
	\end{center}
	\end{subfigure}
	
	\caption{Sampling the neural network at a higher resolution.}
	\label{fig:Resampling}
	\end{center}
\end{figure}

Once the optimization is completed, the global representation of the density field via the NN allows us to sample the field at a finer resolution and  extract a high-resolution topology at no additional cost. This is illustrated in Figure \ref{fig:Resampling}. Observe that  directly smoothing the coarse topology will retain the original topological features. However, sampling at a high resolution can recover  topological features captured by the NN, as can be observed in Figure \ref{fig:Resampling}.

\subsection{Multiple Materials}
\label{sec:multiMaterial}

Next consider maximization of $K$, subject to a mass constraint ($\hat{m} = 0.3$), with two, three and four materials. The material properties are summarized in Table \ref{table:materialProp}.

\begin{table}[ht]
	\centering
\begin{tabular}{||l|l|l|l|l||}
			\hline
			Material & Color code & $E$  & $\nu$ & $\lambda$ \\ \hline
			0 & White & 0.001 & 0.3 & 0  \\ \hline
			1 & Black & 1 & 0.3 & 1  \\ \hline
			2 & Red & 0.2 & 0.3  & 0.2 \\ \hline
			3 & Green & 0.3 & 0.3  & 0.25 \\  \hline
			3 & Blue & 0.4 & 0.3 & 0.3 \\  \hline
\end{tabular}
\caption{Material properties.  }
\label{table:materialProp}
\end{table}

\begin{figure}[H]
\begin{center}
  \begin{subfigure}[b]{.3\textwidth}
  	\begin{center}
    \includegraphics[scale=0.3]{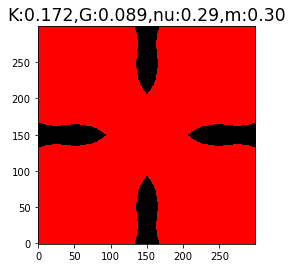}%
		\caption{Two materials.}
	\end{center}
	\end{subfigure}
    \begin{subfigure}[b]{.3\textwidth}
  	\begin{center}
    \includegraphics[scale=0.3]{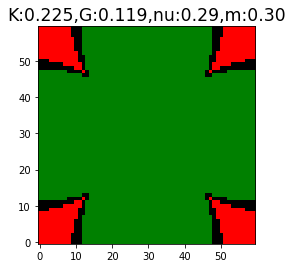}%
		\caption {Three materials.}
	\end{center}
	\end{subfigure}
	\begin{subfigure}[b]{.3\textwidth}
  	\begin{center}
    \includegraphics[scale=0.3]{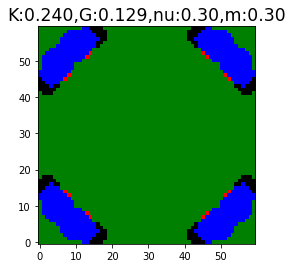}%
		\caption {Four materials.}
	\end{center}
	\end{subfigure}
	\caption{Maximizing $K$ with mass constraint using multiple materials. }
	\label{fig:MultipleMaterials}
	\end{center}
\end{figure}

The results are illustrated in Figure \ref{fig:MultipleMaterials} (compare against Figure \ref{fig:maxBulkModulus} for single material). The final bulk modulus $K^*$ for the three cases are 0.172, 0.225 and 0.240, respectively (compared to 0.091 for single material). Observe that  the bulk modulus increases as additional materials are introduced.

\section{Conclusions}
\label{sec:conclusion}
This paper presents a generalized neural-network-based framework for microstructural optimization where any of the microstructural quantities, namely, bulk, shear, Poisson ratio, volume or mass, can serve as the objective, while the remaining can be subject to constraints.  The error-prone task of sensitivity computation was avoided by exploiting NN's backward propagation. Further, for designing NPR materials, we avoided the use of specialized optimization techniques and heuristics; instead, standard L-BFGS optimization was used. The framework was demonstrated using several numerical experiments. Due to the overhead cost of automatic differentiation (AD), the framework was found to be slower than, say, the MATLAB implementation presented in \cite{xia2015design}. However, we believe the benefits of AD outweigh the computational costs.

There are several directions for future research: extension to multi-physics \cite{das2020multi}, 3D \cite{Andreassen2014}, geometric and material non-linearity \cite{wallin2020nonlinear}, inclusion of stress constraints \cite{collet2018topology}, multi-stable materials \cite{yang20201d}, and inclusion of manufacturing constraints \cite{du2018connecting}.

\section*{Acknowledgments}
\label{sec:acknowledgements}

The authors would like to thank the support of the  National Science Foundation through grant CMMI 1561899, and  the U. S. Office of Naval Research under PANTHER award number N00014-21-1-2916  through Dr. Timothy Bentley.

\section*{Compliance with ethical standards}
\label{sec:ethics}
The authors declare that they have no conflict of interest.

\section*{Replication of Results}
\label{sec:replic}
The Python code pertinent to this paper is available at \href{https://github.com/UW-ERSL/MicroTOuNN}{https://github.com/UW-ERSL/MicroTOuNN}.


\bibliography{allBib}

\end{document}